\documentclass[preprint,prd,superscriptaddress]{revtex4}
\usepackage{amsmath,amssymb}
\usepackage{graphicx}
\usepackage{dcolumn}
\usepackage{bm}
\newcommand{\p}{\partial}
\newcommand{\pslash}{p\kern-1ex /}
\newcommand{\lslash}{l\kern-1ex /}
\newcommand{\kslash}{k\kern-1ex /}
\newcommand{\dslash}{\p\kern-1.2ex /}
\newcommand{\Dslash}{{\cal D}\kern-1.5ex /}
\newcommand{\Aslash}{A\kern-1.2ex /}

\newcommand{\tr}{{\rm tr}}

\newcommand{\Dodwf}{\mathcal{D}}
\newcommand{\bea}{\begin{eqnarray}}
\newcommand{\eea}{\end{eqnarray}}

\newcommand{\BAN}{\begin{eqnarray*}}
\newcommand{\EAN}{\end{eqnarray*}}
\begin{document}

\newcommand{\NTU}{
  Physics Department, 
  National Taiwan University, Taipei~10617, Taiwan  
}

\newcommand{\CQSE}{
  Center for Quantum Science and Engineering,  
  National Taiwan University, Taipei~10617, Taiwan  
}

\newcommand{\CTS}{
  Center for Theoretical Sciences,  
  National Taiwan University, Taipei~10617, Taiwan  
}

\newcommand{\RCAS}{
  Research Center for Applied Sciences, Academia Sinica,
  Taipei~115, Taiwan
}

\preprint{NTUTH-11-505E}
 
\title{Pseudoscalar Meson in Two Flavors QCD 
       with the Optimal Domain-Wall Fermion}


\author{Ting-Wai~Chiu}
\affiliation{\NTU}
\affiliation{\CQSE}
\affiliation{\CTS}


\author{Tung-Han~Hsieh}
\affiliation{\RCAS}


\author{Yao-Yuan~Mao}
\affiliation{\NTU}

\collaboration{TWQCD Collaboration}
\noaffiliation


\begin{abstract}

We perform hybrid Monte Carlo (HMC) simulations of two flavors QCD with the 
optimal domain-wall fermion (ODWF), on the $ 16^3 \times 32 $ 
lattice (with lattice spacing $ a \sim 0.1 $~fm),   
for eight sea-quark masses corresponding to pion masses in the range 228-565 MeV.  
We calculate the mass and the decay constant of the pseudoscalar meson, 
and compare our data with the chiral perturbation theory (ChPT). 
We find that our data is in good agreement 
with the sea-quark mass dependence predicted by 
the next-to-leading order (NLO) ChPT, 
and provides a determination of the low-energy constants
$ \bar{l}_3 $ and $ \bar{l}_4 $,  
the pion decay constant, the chiral condensate, and the
average up and down quark mass.

\end{abstract}

\maketitle


Lattice QCD with exact chiral symmetry \cite{Kaplan:1992bt,Neuberger:1997fp}
is an ideal theoretical framework to study
the nonperturbative physics from the first principles of QCD.
However, it is rather nontrivial to perform Monte Carlo simulation 
such that the chiral symmetry is preserved at a high precision 
and all topological sectors are sampled ergodically. 

Since 2009, TWQCD collaboration has been 
using a GPU cluster (currently constituting of 300 Nvidia GPUs) 
to simulate unquenched lattice QCD 
with the optimal domain-wall fermion (ODWF) \cite{Chiu:2002ir,Chiu:2009wh}.
Mathematically, ODWF is a theoretical framework which preserves 
the chiral symmetry optimally with a set of analytical weights, 
$ \{ \omega_s, s = 1, \cdots, N_s \} $, 
one for each layer in the fifth dimension \cite{Chiu:2002ir}. 
Thus the artifacts due to the chiral
symmetry breaking with finite $ N_s $ can be reduced to the minimum, 
especially in the chiral regime.
The 4-dimensional effective Dirac operator of massless ODWF is
\BAN
\begin{aligned}
D &= m_0 [1+ \gamma_5 S_{opt}(H_w) ], \\
S_{opt}(H_w) &= \frac{1-\prod_{s=1}^{N_s} T_s}{1 + \prod_{s=1}^{N_s} T_s}, \quad 
T_s = \frac{1-\omega_s H_w}{1+\omega_s H_w},  
\end{aligned}
\EAN
which is exactly equal to the Zolotarev optimal rational approximation 
of the overlap Dirac operator. That is,   
$ S_{opt}(H_w) = H_w R_Z(H_w) $, where $ R_Z(H_w)$ is the optimal 
rational approximation of $ (H_w^2)^{-1/2} $ 
\cite{Akhiezer:1992, Chiu:2002eh}.

Recently we have demonstrated that it is feasible to perform a large-scale unquenched QCD 
simulation which not only preserves the chiral symmetry to a good precision, 
but also samples all topological sectors ergodically \cite{Chiu:2011dz}.   
To recap, we perform HMC simulations of 2 flavors QCD on a $ 16^3 \times 32 $ lattice, 
with ODWF at $ N_s = 16 $, and plaquette gauge action at $ \beta = 5.95 $. 
Then we compute the low-lying eigenmodes of the overlap Dirac operator, 
and use its index to obtain the topological charge of each 
gauge configuration, and from which we compute the topological susceptibility
for 8 sea-quark masses, each of 300 configurations. 
Our result of the topological susceptibility agrees with the 
sea-quark mass dependence predicted by
the NLO ChPT \cite{Mao:2009sy}, 
and provides the first determination of both
the pion decay constant and the chiral condensate 
simultaneously from the topological susceptibility. 

In this paper, we perform further simulations and increase the ensemble 
of each sea-quark mass from 300 to 500 configurations. That is, for each 
sea-quark mass, we generate 5000 trajectories after thermalization, 
and sample one configuration every 10 trajectories. 
Then we compute the valence quark propagators and 
the time-correlation function of the pseudoscalar meson operator, 
and from which we extract the mass $ M_\pi $ and the decay constant $ F_\pi $ 
of the pseudoscalar meson.  
We compare our results of $ M_\pi $ and $ F_\pi $ with the 
NLO ChPT \cite{Gasser:1984gg}, and  
find that our results are in good agreement with 
the sea-quark mass dependence predicted by NLO ChPT, 
and from which we obtain the low-energy constants 
$ F $, $ \Sigma $, $ \bar{l}_3 $ and $ \bar{l}_4 $.
With the low-energy constants, we determine the average up and down quark mass 
$m_{ud}^{\overline{\rm MS}}(\mathrm{2~GeV})$, and the chiral condensate
$\Sigma^{\overline{\rm MS}}(\mathrm{2~GeV})$.


First, we outline our HMC simulation of 2 flavors QCD with ODWF. 
Starting from the ODWF action $ S = \bar\Psi \Dodwf \Psi $ \cite{Chiu:2002ir} 
on the 5D lattice,    
we separate the even and the odd sites 
(the so-called even-odd preconditioning)
on the 4D lattice,  
and rewrite $ \Dodwf $ as
\BAN
\Dodwf(m_q)=
S_1^{-1}
\begin{pmatrix}
1 & 0 \\
M_5 D_w^{\text{OE}} & 1
\end{pmatrix}
\begin{pmatrix}
1 & 0 \\
0 & C
\end{pmatrix}
\begin{pmatrix}
1 & M_5 D_w^{\text{EO}} \\
0 & 1
\end{pmatrix}
S_2^{-1}, 
\label{eq:D_odwf_decomp}
\EAN
where $ m_q $ denotes the bare quark mass,   
$D_w$ denotes the standard Wilson Dirac operator plus a negative parameter $-m_0 $ 
(Here $ m_0 = 1.3 $ in this paper.),
and $D_w^{\text{EO/OE}}$ denotes the part of $ D_w $ with gauge links pointing 
from odd/even sites to even/odd sites, and 
\BAN
\begin{aligned}
\label{eq:m5}
M_5 &= \left[(4-m_0) + {\omega}^{-1/2}(1-L)(1+L)^{-1}{\omega}^{-1/2}\right]^{-1}, \\
(\omega)_{ss'} &= \omega_s \delta_{ss'},  \\
L = P_+ & L_+ + P_- L_-, \quad P_\pm = (1\pm \gamma_5)/2, \quad L_-=(L_+)^T, \\ 
(L_+)_{ss'} &= \left\{ 
    \begin{array}{ll} \delta_{s-1,s'}, & 1 < s \leq N_s \\ 
        -(m_q/2m_0) \delta_{N_s,s'}, & s = 1 \end{array}\right.;  \\  
S_1 &= M_5 {\omega}^{-1/2}, \quad S_2= (1 +  L)^{-1} {\omega}^{-1/2}, \\ 
\label{eq:c_def}
C &= 1 - M_5 D_w^{\text{OE}} M_5 D_w^{\text{EO}}. 
\end{aligned}
\EAN
Since $ \det\Dodwf = \det S_1^{-1} \cdot \det C \cdot \det S_2^{-1} $, and
$ S_1 $ and $ S_2 $ do not depend on the gauge field, we can just use $ C $
for the HMC simulation. After including the Pauli-Villars fields (with $ m_q = 2 m_0 $), 
the pseudo-fermion action for 2 flavors QCD ($ m_u = m_d $) can be written as
\bea
\label{eq:Spf}
S_{pf} = \phi^\dagger C_{PV}^\dagger ( C C^\dagger)^{-1} C_{PV} \phi, \quad C_{PV} \equiv C(2m_0).
\eea

In the HMC simulation \cite{Duane:1987de}, 
we first generate random noise vector $ \xi $ with Gaussian distribution,
then we obtain $ \phi = C_{PV}^{-1} C \xi $ using the conjugate gradient (CG).
With fixed $ \phi $, the system is evolved under a fictituous Hamiltonian dynamics,
the so-called molecular dynamics (MD). In the MD, we use the Omelyan integrator \cite{Takaishi:2005tz},
and the Sexton-Weingarten multiple-time scale method \cite{Sexton:1992nu}.
The most time-consuming part in the MD is to compute the vector $ \eta = (C C^\dagger)^{-1} C_{PV} \phi $
with CG, which is required for the evaluation of the fermion force in the equation
of motion for the conjugate momentum of the gauge field. Here we take advantage of 
the remarkable floating-point capability of the Nvidia GPU, and perform the 
CG with mixed precision \cite{Chiu:2011rc}.
Moreover, the computations of the gauge force and the fermion force, and the update of the gauge field
are also ported to the GPU. In other words, almost the entire HMC simulation is 
performed within a single GPU.

Furthermore, we introduce an auxillary heavy fermion field with mass $ m_H $ ($ m_q \ll m_H < 2 m_0 $), 
similar to the case of the Wilson fermion \cite{Hasenbusch:2001ne}.
For two flavors QCD, the pseudofermion action (with $ C_H \equiv C(m_H) $) becomes,  
\BAN
S_{pf}^H = \phi^{\dagger} C_H^{\dagger} (CC^{\dagger})^{-1} C_H \phi + 
           \phi_H^{\dagger} C_{PV}^{\dagger} (C_H C_H^{\dagger})^{-1} C_{PV} \phi_H, 
\EAN
which gives exactly the same fermion determinant of (\ref{eq:Spf}).   
Nevertheless, the presence of the heavy fermion field plays a crucial role in 
reducing the light fermion force and its fluctuation, thus diminishes the change 
of the Hamiltonian in the MD trajactory, and enhances the acceptance rate.
A detailed description of our HMC simulations will be 
presented in a forthcoming paper \cite{Chiu:HMC}.


We determine the lattice spacing by heavy quark potential 
which is extracted from Wilson loops 
of size $(R_1,R_2,T)$, where $R_1$,  
$R_2$ and $ T $ are the sizes in spatial and temporal directions. 
The spatial distance between the heavy quark and antiquark is 
$R = \sqrt{R_1^2+R_2^2} $. We measure all planar and non-planar
Wilson loops $ W $ with $a \le R \le 8a$ and $a \le T \le 8a$. 
Fitting the data of $ W(R,T) $ to the formula
$ \left< W \right> = C \exp(-T\, V(R)) $, we 
obtain the heavy quark potential $V(R)$ as a function of $R$.
Here we have used all 5000 trajectories after thermalization, 
and we estimate the error of $ V(R) $ using the jackknife method
with the bin size of which the statistical error saturates.
Then we fit our data of $V$ to the formula
\bea
\label{eq:V}
V(R) = A + \frac{B}{R} + \sigma R, 
\eea 
to obtain $ A $, $ B $, and $ \sigma $.
We summarize our results in Table \ref{tab:ABsigma_b595}. 

\begin{table}[th]
\begin{center}
\begin{tabular}{|c|cccc|c|}
\hline
$ m_q a $ & $ A $ & $ B $ & $ \sigma $ & $ \chi^2$/dof & $a$[fm] \\
\hline
0.01  &  0.7777(57)  &  -0.3814(70) &  0.0577(10)  &  0.0329  &  0.1045(13)  \\
0.02  &  0.7827(46)  &  -0.3818(41) &  0.0584(9)   &  0.0275  &  0.1051(10)  \\ 
0.03  &  0.7792(54)  &  -0.3789(62) &  0.0595(9)   &  0.0368  &  0.1060(12)  \\
0.04  &  0.7916(71)  &  -0.3995(78) &  0.0598(13)  &  0.0440  &  0.1071(16)  \\
0.05  &  0.7797(73)  &  -0.3798(72) &  0.0615(13)  &  0.0456  &  0.1078(16)  \\
0.06  &  0.7762(50)  &  -0.3785(44) &  0.0628(11)  &  0.0458  &  0.1089(11)  \\
0.07  &  0.7783(47)  &  -0.3855(53) &  0.0633(8)   &  0.0255  &  0.1097(10)  \\
0.08  &  0.7719(69)  &  -0.3744(64) &  0.0649(12)  &  0.0569  &  0.1105(14)  \\
\hline
\end{tabular}
\end{center}
\caption{
The parameters of $ A $, $ B $, and $ \sigma $ obtained by fitting our data of 
heavy quark potential $ V(R) $ to Eq. (\ref{eq:V}), together with the $ \chi^2$/dof 
of the fit. The lattice spacing in the last column is obtained by (\ref{eq:a}). 
}
\label{tab:ABsigma_b595}
\end{table}


Using the empirical formula deduced by Sommer \cite{Sommer:1993ce}, 
\bea
F(r_0)r_0^2 = 1.65, \hspace{5mm}  F(r) \equiv \frac{d}{dr}V(r) = -\frac{B}{r^2}+\sigma, 
\eea
and setting the Sommer parameter $r_0 = 0.49$~fm, we obtain the lattice spacing  
\bea
\label{eq:a}
a = r_0 \sqrt{\frac{\sigma}{1.65+B}}, 
\eea
where the results are given in the last column of Table \ref{tab:ABsigma_b595}. 
Using the linear fit, we obtain the lattice spacing in the chiral limit,
$ a = 0.1034(1)(2) $ fm with $ \chi^2$/dof = 0.10, where the systematic error
is estimated by varying the number of sea-quark masses. 
This gives the inverse lattice spacing $ a^{-1} = 1.908(2)(4) $~GeV. 
 
We compute the valence quark propagator of the 4D effective Dirac operator 
with the point source at the origin, and with parameters exactly the same 
as those of the sea-quarks.
First, we solve the following linear system (with even-odd preconditioned CG), 
\bea
\label{eq:DY}
{\cal D}(m_q) |Y \rangle = {\cal D}(2m_0) B^{-1} |\mbox{source vector} \rangle, 
\eea
where $ B^{-1}_{x,s;x',s'} = \delta_{x,x'}(P_{-}\delta_{s,s'}+P_{+}\delta_{s+1,s'}) $
with periodic boundary conditions in the fifth dimension.
Then the solution of (\ref{eq:DY}) gives the valence quark propagator  
\BAN
\label{eq:v_quark}
(D_c + m_q)^{-1}_{x,x'} = \left( 2 m_0 - m_q \right)^{-1} \left[ (BY)_{x,1;x',1} - \delta_{x,x'} \right]. 
\EAN

To measure the chiral symmetry breaking due to finite $N_s$, we compute the residual mass with 
the formula \cite{Chen:2012jy}
\bea
\begin{aligned}
m_{res} 
&= \left< \frac{ \tr(D_c + m_q)^{-1}_{0,0} }{ \tr[(D_c^\dagger + m_q)(D_c+m_q)]^{-1}_{0,0} } \right>_{\{U\}} - m_q,  
\label{eq:mres}
\end{aligned}
\eea
where 
$ (D_c + m_q)^{-1} $ denotes the valence quark propagator with $ m_q $ equal to the sea-quark mass, 
tr denotes the trace running over the color and Dirac indices, and the subscript $ \{U\} $ denotes averaging 
over an ensemble of gauge configurations. 
In Table \ref{tab:mres_b595}, we list the residual masses for eight sea quark masses, 
together with those obtained by setting $ \omega_s = 1 $ (polar approximation of the sign 
function of $ H_w $) in the valance quark propagator. In the latter case, even though the 
chiral symmetry of the valence quarks is different from that of the sea quarks, it
may serve as an estimate of the residual mass in the unitary limit with $ \omega_s = 1 $.    
We see that turning on $ \{ \omega_s \} $ with $ \lambda_{min}/\lambda_{max} = 0.02/6.40 $, 
the residual mass is decreased by a factor of 25-40, while the cost of computing quark propagators 
is increased by a factor of 2-5. Moreover, for $ m_q a = 0.01 $, we also computed the residual mass with $ N_s = 32 $ 
and $ \omega_s = 1 $, and obtained $ m_{res} = 0.002746(13) $ which is 6 times larger than  
that of turning on $ \{ \omega_s \} $ with $ N_s = 16 $ and $ \lambda_{min}/\lambda_{max} = 0.02/6.40 $, 
while the cost is almost the same in both cases.   
This suggests that ODWF is a viable way to preserve the chiral symmetry 
on the lattice, without increasing $ N_s $. 
For ODWF, using the linear fit, we obtain the residual mass in the chiral limit, 
$ m_{res} a = 0.00040(4) $, less than $ 5\% $ of the lightest sea quark mass.
In the following, it is understood that each bare sea-quark mass $ m_q $ is
corrected by its residual mass, i.e., $ m_q \rightarrow m_q + m_{res} $.

\begin{table}[th]
\begin{center}
\begin{tabular}{|c|c|c|c|}
\hline
$ m_q a $ & $ m_{res} $(ODWF) & $ m_{res} ( \omega_s = 1) $ & ratio \\
\hline
0.01  &  0.000418(31)  & 0.01064(17)  &  0.039(3)  \\
0.02  &  0.000380(29)  & 0.01139(15)  &  0.033(3)  \\ 
0.03  &  0.000269(40)  & 0.01047(13)  &  0.026(4)  \\
0.04  &  0.000259(43)  & 0.01043(12)  &  0.025(4)  \\
0.05  &  0.000269(41)  & 0.01000(13)  &  0.027(4)  \\
0.06  &  0.000357(47)  & 0.01029(11)  &  0.035(4)  \\
0.07  &  0.000248(45)  & 0.00988(15)  &  0.025(6)  \\
0.08  &  0.000219(38)  & 0.00991(13)  &  0.022(4)  \\
\hline
\end{tabular}
\end{center}
\caption{
The residual mass (second column) versus the sea quark mass for two flavors QCD with ODWF.
The third column is the residual mass obtained by setting $ \omega_s = 1$ 
in the valence quark propagator. The last column is the ratio $ m_{res} $(ODWF)/$ m_{res}(\omega_s = 1) $. 
}
\label{tab:mres_b595}
\end{table}



\begin{figure*}[tb]
\begin{center}
\begin{tabular}{@{}c@{}c@{}}
\includegraphics*[width=7.5cm,clip=true]{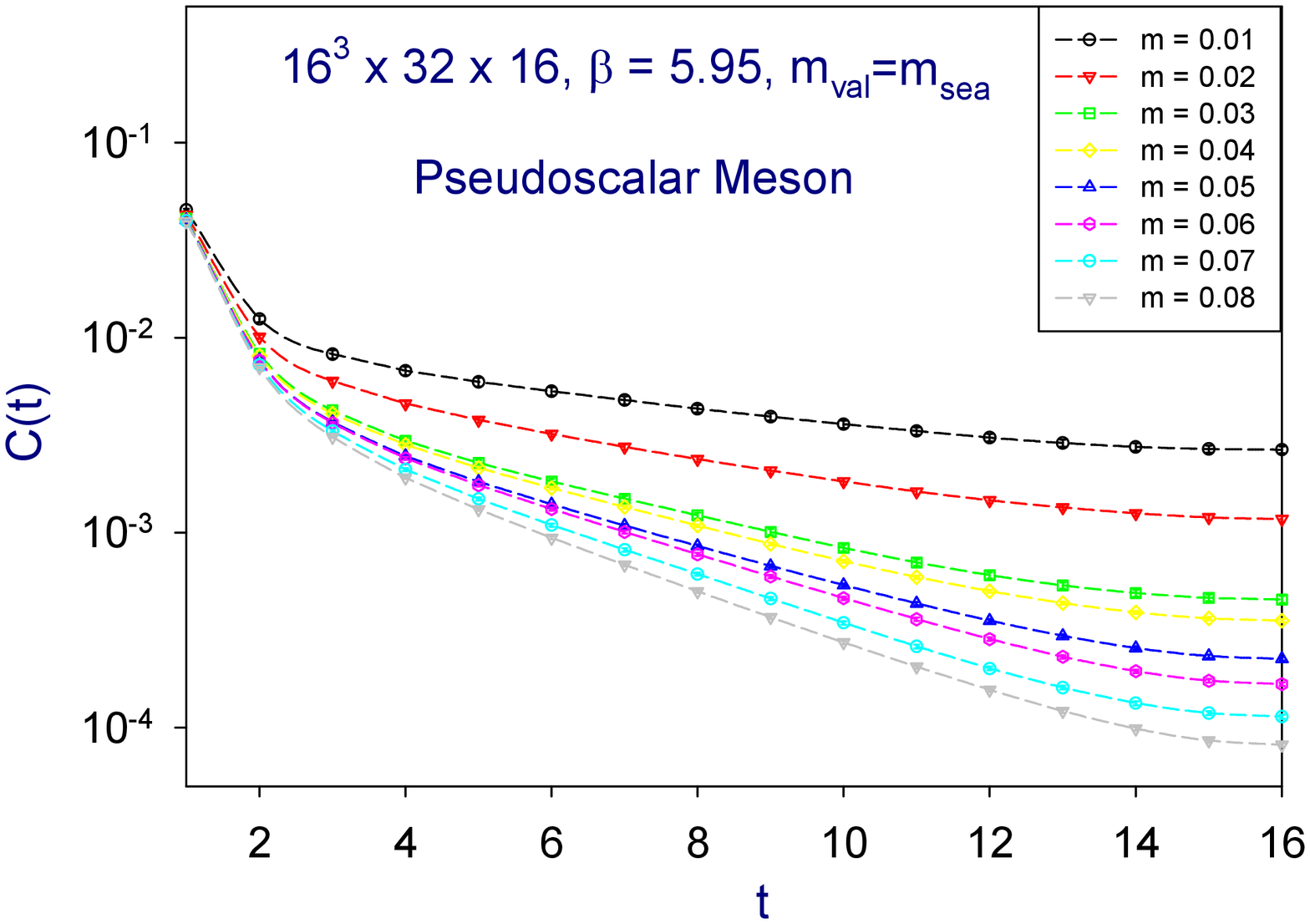}
&
\includegraphics*[width=7.5cm,clip=true]{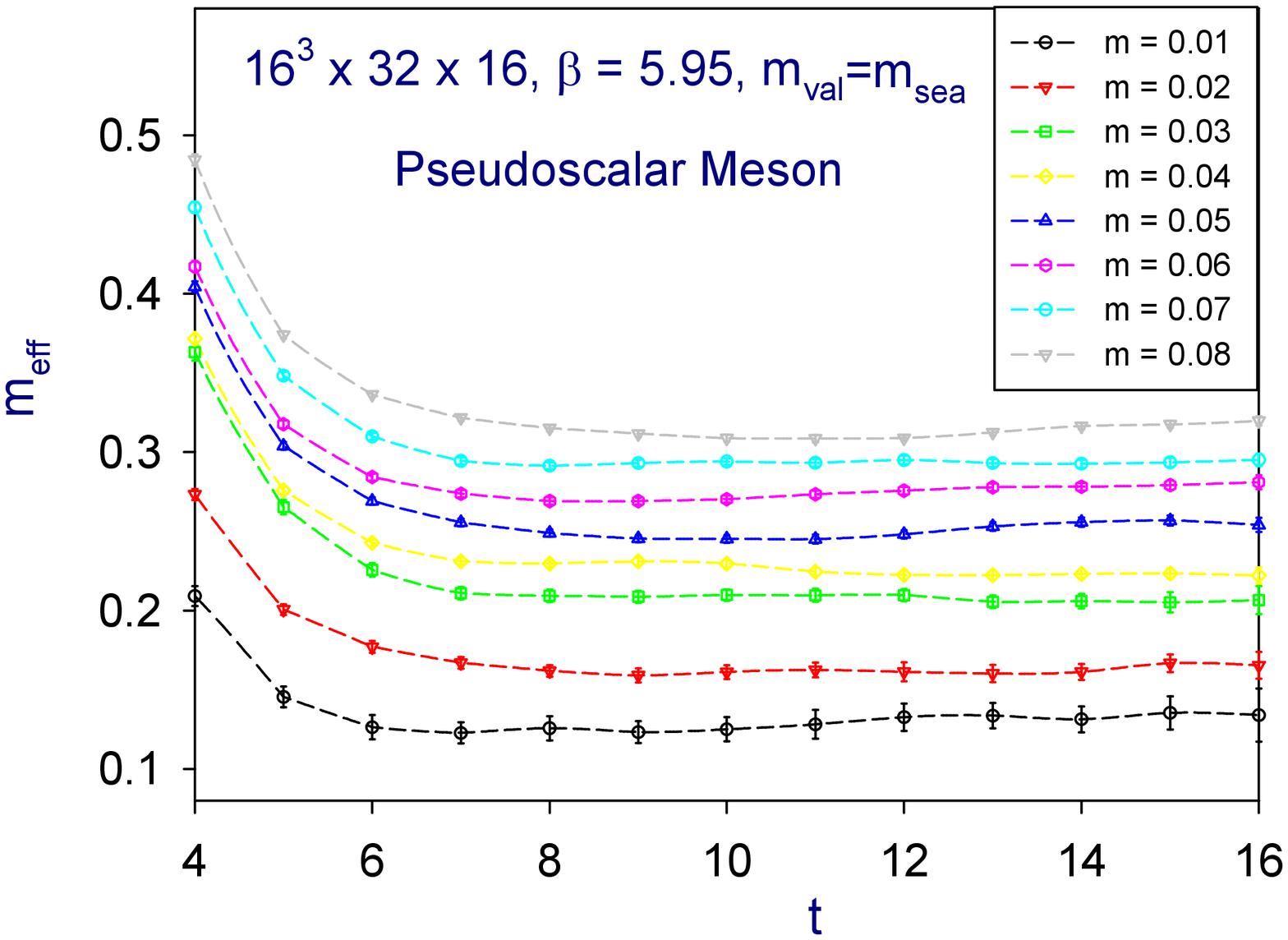}
\\ (a) & (b)
\end{tabular}
\caption{(color online) (a) The time-correlation function of the pseudoscalar meson for eight sea quark masses. 
                        (b) The effective mass of (a). 
         The dashed line connecting the data points of the same sea-quark mass is for guiding the eyes.}
\label{fig:G55a_meff_b595_nf2}
\end{center}
\end{figure*}

Using the valence quark propagator with quark mass equal to the sea-quark mass, 
we compute the time-correlation function of the pseudoscalar interpolator
\BAN
\label{eq:Ct}
C(t) = \sum_{ \vec{x} } \tr \{ \gamma_5 ( D_c + m_q )^{-1}_{0,x} \gamma_5
                ( D_c + m_q )^{-1}_{x,0} \},
\EAN
where the trace runs over the Dirac and color space.
In Fig. \ref{fig:G55a_meff_b595_nf2}, we plot  
$ C(t) $ and its effective mass 
$ m_{\rm eff}(t) = \cosh^{-1} \{ [ C(t+1) + C(t-1) ]/(2 C(t)) \} $
for eight sea quark masses respectively.  
Then $ \langle C(t) \rangle $ is fitted to the formula
$ Z [ e^{-M_{\pi} t} + e^{-M_{\pi} (T-t)} ]/(2 M_{\pi} ) $
to extract the pion mass $ M_{\pi} $ and the decay constant
$ F_{\pi} = m_q \sqrt{2Z}/M_{\pi}^2 $, 
where the excited states have been neglected.
Here we have chosen the fitting range $ [t_1, t_2] $ in which 
the effective mass attaining a plateau,  
and we estimate the errors of $ M_{\pi} $ and $ F_{\pi} $ 
using the jackknife method with the bin size of 15 configurations
of which the statistical error saturates.

\begin{figure*}[tb]
\begin{center}
\begin{tabular}{@{}c@{}c@{}}
\includegraphics*[width=7.5cm,clip=true]{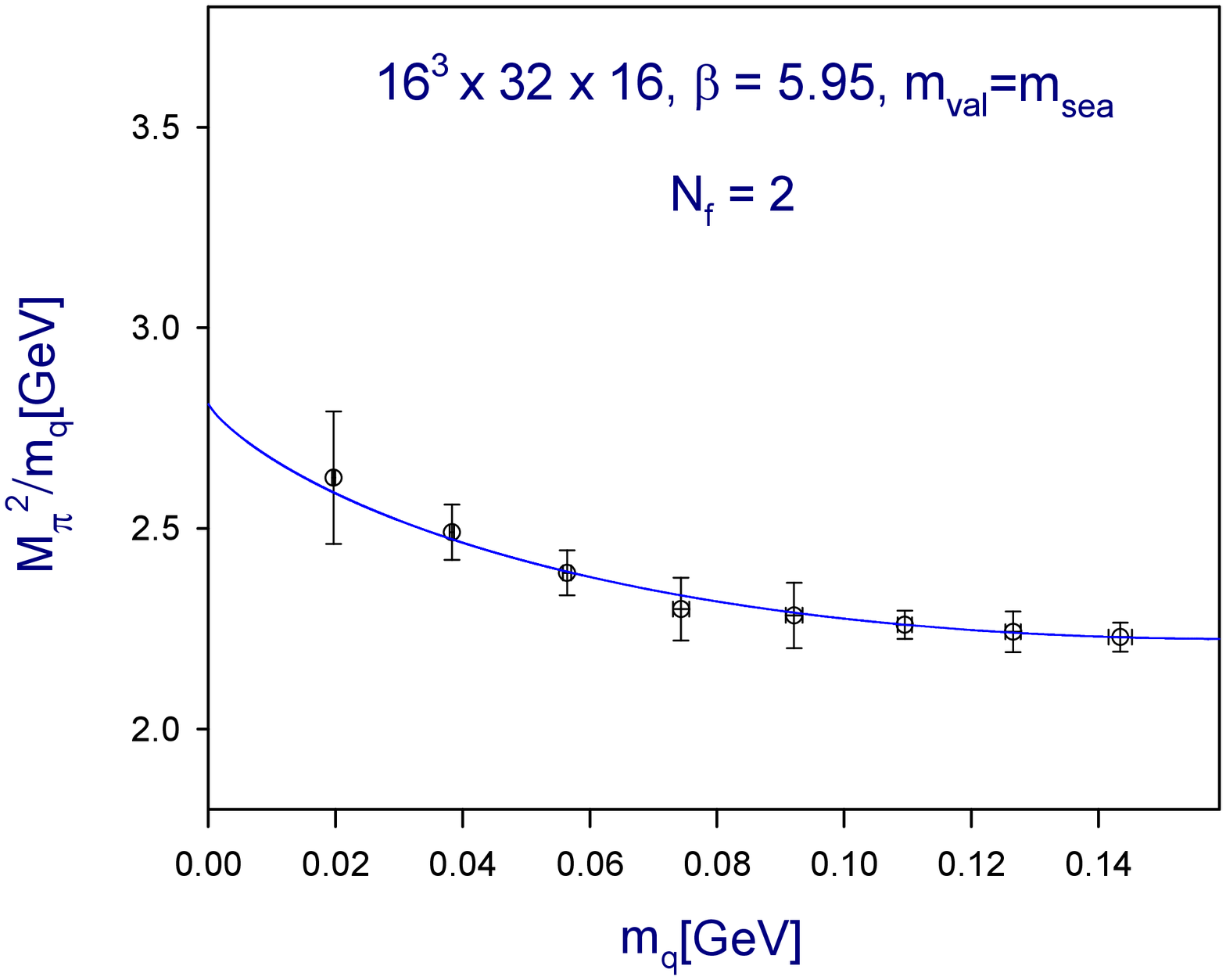}
&
\includegraphics*[width=7.5cm,clip=true]{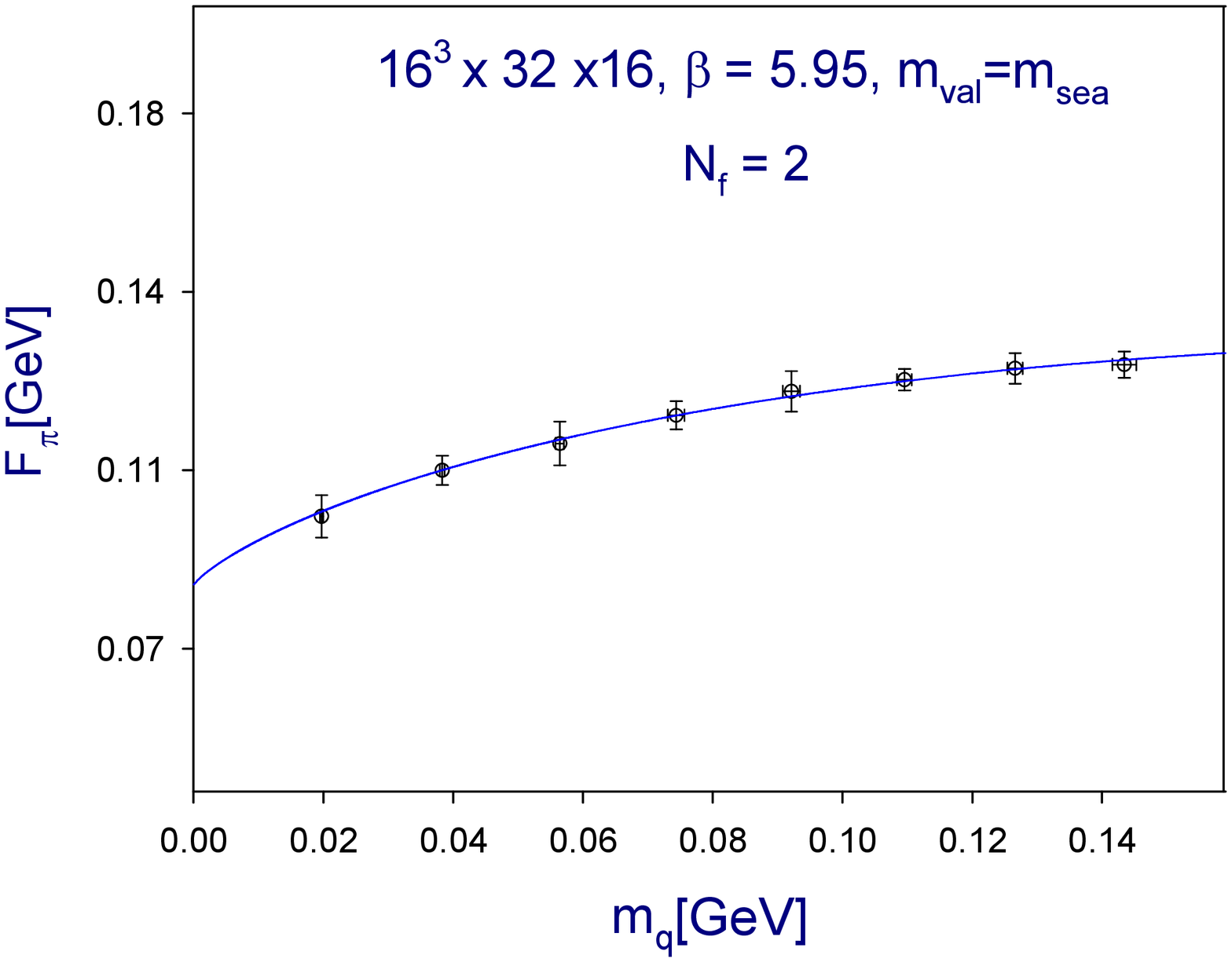}
\\ (a) & (b)
\end{tabular}
\caption{Physical results of 2 flavors QCD with ODWF 
         (a) $ M_\pi^2/m_q $, and (b) $ F_\pi $.
          The solid lines are the simultaneous fits to the NLO ChPT, for seven sea-quark masses ($m_q a = 0.02-0.08$).
          Note that the data points of the lightest pion are also falling on the curves of NLO ChPT fit. 
}
\label{fig:Mpi2omq_Fpi_b595_nf2}
\end{center}
\end{figure*}

We make the correction for the finite volume effect 
using the estimate within ChPT calculated up 
to $ {\cal O}(M_\pi^4/(4 \pi F_\pi)^4 ) $ \cite{Colangelo:2005gd}.
In Table \ref{tab:Mpi_Fpi_b595_nf2}, we give the values of $ M_\pi $
and $ F_\pi $ (with finite volume corrections), together with
their finite volume correction factors computed using the formulas
given in \cite{Colangelo:2005gd}.
In Fig. \ref{fig:Mpi2omq_Fpi_b595_nf2},  
we plot $ M_\pi^2 /m_q $ and $ F_\pi $ 
versus $ m_q $ respectively.
For the lighest pion, $ M_\pi L \simeq 2.0 $, 
the formulas for finite volume correction may be unreliable, 
according to Ref. \cite{Colangelo:2005gd}. 
Thus, we perform the ChPT fit with the lightest pion excluded.
Then we will check whether the lightest pion falls on the curve of the ChPT fit.

\begin{table}[th]
\begin{center}
\begin{tabular}{|c|cccccc|}
\hline
$ m_q a $ & $ [t_1, t_2] $ & $\chi^2$/dof & $ M_\pi$[GeV] & $ F_\pi$[GeV] & $ 1+R_{M_\pi} $ & $ 1+R_{F_\pi} $   \\
\hline
0.01  &  [8,13] & 1.04 & 0.2275(76) & 0.0970(42) &  1.0815 &   0.7940   \\
0.02  &  [9,14] & 0.60 & 0.3089(49) & 0.1060(29) &  1.0301 &   0.9271   \\
0.03  &  [6,13] & 0.53 & 0.3672(56) & 0.1114(44) &  1.0158 &   0.9629   \\
0.04  &  [6,13] & 0.84 & 0.4135(93) & 0.1170(28) &  1.0091 &   0.9789   \\
0.05  &  [7,13] & 0.41 & 0.4586(100) & 0.1217(40) &  1.0055 &   0.9874   \\
0.06  &  [7,12] & 1.21 & 0.4976(59) & 0.1240(21) &  1.0037 &   0.9918   \\
0.07  &  [9,13] & 0.44 & 0.5327(74) & 0.1263(30) &  1.0026 &   0.9943   \\
0.08  &  [6,15] & 0.88 & 0.5654(78) & 0.1270(26) &  1.0020 &   0.9959   \\
\hline
\end{tabular}
\end{center}
\caption{
Summary of the data of $ M_\pi $ and $ F_\pi $. 
The second column is the range $ [t_1, t_2] $
of the time-correlation function used for fitting,  
the third column is the $ \chi^2$/dof of the fit, 
and the last two columns are finite volume corrections
for $ M_\pi $ and $ F_\pi $ respectively. 
}
\label{tab:Mpi_Fpi_b595_nf2}
\end{table}

Taking into account of the correlation between $ M_\pi^2/m_q $ and $ F_\pi $ for the same sea-quark mass,  
we fit our data to the formulas of NLO ChPT \cite{Gasser:1984gg}
\bea
\label{eq:mpi2omq_NLO_Nf2}
\frac{M_\pi^2}{m_q} &=& \frac{2 \Sigma}{F^2}  \left[ 1 
+ \left(\frac{\Sigma m_q }{16 \pi^2 F^4}\right) \ln\left(\frac{2 \Sigma m_q}{F^2 \Lambda_3^2} \right) \right], \\
\label{eq:fpi_NLO_Nf2}
F_\pi &=& F \left[ 1 -  \left(\frac{\Sigma m_q}{8 \pi^2 F^4 } \right) \ln \left( \frac{2 \Sigma m_q}{ F^2 \Lambda_4^2} \right) \right], 
\eea
where $ \Lambda_3 $ and $ \Lambda_4 $ are related to the low energy constants $ \bar l_3 $ and $ \bar l_4 $ as follows.   
\BAN
\bar l_3 = \ln \left( \frac{\Lambda_3^2}{m_{\pi^{\pm}}^2} \right), \  
\bar l_4 = \ln \left( \frac{\Lambda_4^2}{m_{\pi^{\pm}}^2} \right), \ m_{\pi^{\pm}} = 0.140 \mbox{ GeV}.  
\EAN

The strategy of our data fitting is to search for the values of the parameters 
$ \Sigma $, $ F $, $ \Lambda_3 $ and $ \Lambda_4 $ such that they minimize
\BAN
\label{eq:chi2}
\chi^2 = \sum_{i} V_i^T C_i^{-1} V_i, \ 
V_i = \begin{pmatrix}
(M_\pi^2/m_q)_i -(M_\pi^2/m_q)_i^{\tiny\mbox{ChPT}} \\
(F_\pi)_i - (F_\pi)_i^{\tiny\mbox{ChPT}} 
\end{pmatrix}, 
\EAN
where $ C_i $ is the $ 2 \times 2 $ covariance matrix for $ M_\pi^2/m_q $ and $ F_\pi $ with the same sea-quark mass.  

For seven sea-quark masses corresponding to pion masses in the range $309-565$ MeV, 
our fit gives 
\bea
\label{eq:sigma13_7}
\Sigma &=& [0.2140(13)(11) \mbox{ GeV}]^3, \\
\label{eq:F2_8}
F &=& 0.0835(10)(14) \mbox{ GeV},  \\  
\label{eq:l3_7}
\bar l_3 &=& 4.156(34)(122), \\
\label{eq:l4_7}
\bar l_4 &=& 4.473(36)(46),  
\eea   
with $\chi^2$/dof = 0.07, 
where the systematic errors are estimated by varying the number of data points from 
7 ($ M_\pi \le 565 $ MeV) to 4 ($ M_\pi \le 459 $ MeV). In Fig. (\ref{fig:Mpi2omq_Fpi_b595_nf2}), 
we see that the data points of the lightest pion also fall on the curves of NLO ChPT fit. 
This seems to suggest that the finite volume corrections for the lightest pion (with $ M_\pi L \simeq 2.0 $)
may be correct. 

To obtain the physical bare quark mass, we use the physical ratio 
$ \left( M_\pi/F_\pi \right)^{phys} = 0.135/0.093 = 1.45 $ 
as the input, and solve the equation $ M_\pi(m_q)/F_\pi(m_q) = 1.45 $
to obtain the physical bare quark mass $ m_q^{phys} = 0.00519(15)(18) $~GeV. 
From (\ref{eq:fpi_NLO_Nf2}) and (\ref{eq:mpi2omq_NLO_Nf2}),  
we obtain the pion decay constant and the pion mass 
at the physical point,  
\bea
\label{eq:Fpi_phys}  
F_\pi &=& 0.0898(12)(14) \mbox{ GeV}, \\ 
\label{eq:Mpi_phys}  
M_\pi &=& 0.1298(50)(55) \mbox{ GeV}.   
\eea 
Since we have used the physical ratio 1.45
as the input, in principle, we can only regard 
either (\ref{eq:Fpi_phys}) or (\ref{eq:Mpi_phys}) 
as our predicted physical result.

In order to convert the chiral condensate $ \Sigma $ and 
the average $ m_u $ and $ m_d $ to those 
in the $\overline{\mathrm{MS}}$ scheme, 
we calculate the renormalization factor 
$Z_s^{\overline{\mathrm{MS}}}(\mathrm{2~GeV}) $ 
using the non-perturbative renormalization technique
through the RI/MOM scheme \cite{Martinelli:1994ty}, 
and our result is \cite{Chiu:2011np}
\bea
\label{eq:Zs}
Z_s^{\overline{\mathrm{MS}}}(\mathrm{2~GeV}) = 1.244(18)(39).  
\eea
Then the values of $ \Sigma $ and the average of $ m_u $ and $ m_d $ are transcribed to
\bea
\label{eq:sigmaMS}
\Sigma^{\overline{{\mathrm{MS}}}}(\mbox{2 GeV}) &=& [230(4)(6) \mbox{ MeV}]^3,  \\
\label{eq:mudMS}
m_{ud}^{\overline{{\mathrm{MS}}}}(\mbox{2 GeV}) &=& 4.17(13)(19) \mbox{ MeV}, 
\eea
where the systematic errors follow from those in Eqs. (\ref{eq:sigma13_7}) and (\ref{eq:Zs}). 

Since our calculation is done at a single lattice spacing 
the discretization error cannot be quantified reliably, but
we do not expect much larger error because our lattice
action is free from $O(a)$ discretization effects. 

We also investigated to what extent our results of the low-energy constants 
depending on the chiral symmetry of the valence quark propagators.
We repeated above analysis with valence quark propagators computed 
with $N_s = 32$ and $\lambda_{min}/\lambda_{max} = 0.01/6.4 $, 
which has the residual mass $ m_{res} a = 0.000191(12) $ in the chiral limit.
The low-energy constants turn out to be in good agreement with those in 
(\ref{eq:sigma13_7})-(\ref{eq:l4_7}). 

Moreover, our present results 
of the chiral condensate (\ref{eq:sigmaMS}) and 
the pion decay constant (\ref{eq:Fpi_phys})
are consistent with our recent results 
extracted from the topological susceptibility \cite{Chiu:2011dz}. 
%

In general, our results of the $ SU(2) $ low-energy constants, 
the chiral condensate, and the average up and down quark mass  
are compatible with those obtained by other lattice groups 
using unitary dynamical quarks with $ N_f=2 $, e.g., 
Ref. \cite{Noaki:2008iy}. 
A detailed comparison with all lattice results
\cite{Colangelo:2010et} is beyond the scope of this paper. 
 
To conclude, our results of the mass and the decay constant
of the pseudoscalar meson are in good agreement 
with the sea-quark mass dependence predicted by 
the next-to-leading order (NLO) ChPT,   
and provide a determination of the low-energy constants
$ \bar{l}_3 $ and $ \bar{l}_4 $,  
the pion decay constant, the chiral condensate, and the
average up and down quark mass.
Together with our recent result of the topological susceptibility \cite{Chiu:2011dz},  
these suggest that the nonperturbative chiral dynamics of the sea quarks 
are well under control in our HMC simulations.     
Moreover, this study also shows that it is feasible  
to perform large-scale simulations of unquenched lattice QCD, 
which not only preserve the chiral symmetry to a good precision,    
but also sample all topological sectors ergodically.    
This provides a new strategy to tackle QCD nonperturbatively 
from the first principles.

%
  This work is supported in part by the National Science Council
  (Nos. NSC99-2112-M-002-012-MY3, NSC99-2112-M-001-014-MY3) and NTU-CQSE (No. 10R80914-4). 
  We also thank NCHC and NTU-CC for providing facilities to perform part of our calculations. 
    


\begin{thebibliography}{99}

\bibitem{Kaplan:1992bt}
D.~B.~Kaplan,
Phys.\ Lett.\ B {\bf 288}, 342 (1992);
Nucl.\ Phys.\ Proc.\ Suppl.\  {\bf 30}, 597 (1993).

\bibitem{Neuberger:1997fp}
  H.~Neuberger,
  Phys.\ Lett.\ B {\bf 417}, 141 (1998); 
%
%
  R.~Narayanan and H.~Neuberger,
  Nucl.\ Phys.\ B {\bf 443}, 305 (1995).

\bibitem{Chiu:2002ir}
  T.~W.~Chiu,
  Phys.\ Rev.\ Lett.\  {\bf 90}, 071601 (2003);
%
  Phys.\ Lett.\ B {\bf 552}, 97 (2003);
%
  hep-lat/0303008
%

\bibitem{Chiu:2009wh}
  T.W.~Chiu {\it et al.} [TWQCD Collaboration],
  PoS {\bf LATTICE2009}, 034 (2009).
  [arXiv:0911.5029 [hep-lat]]

\bibitem{Akhiezer:1992}
N.~I.~Akhiezer,
"Theory of approximation", Reprint of 1956 English translation, Dover,
New York, 1992.

\bibitem{Chiu:2002eh}
  T.~W.~Chiu, T.~H.~Hsieh, C.~H.~Huang and T.~R.~Huang,
  Phys.\ Rev.\  D {\bf 66}, 114502 (2002).

\bibitem{Chiu:2011dz}
  T.~W.~Chiu, T.~H.~Hsieh and Y.~Y.~Mao,
  Phys.\ Lett.\  B {\bf 702}, 131 (2011).
  [arXiv:1105.4414 [hep-lat]]

\bibitem{Mao:2009sy}
  Y.~Y.~Mao and T.~W.~Chiu  [TWQCD Collaboration],
  Phys.\ Rev.\  D {\bf 80}, 034502 (2009).

\bibitem{Gasser:1984gg}
  J.~Gasser and H.~Leutwyler,
  Nucl.\ Phys.\  B {\bf 250}, 465 (1985).

\bibitem{Duane:1987de}
  S.~Duane, A.~D.~Kennedy, B.~J.~Pendleton and D.~Roweth,
  Phys.\ Lett.\  B {\bf 195}, 216 (1987).

\bibitem{Takaishi:2005tz}
  T.~Takaishi and P.~de Forcrand,
  Phys.\ Rev.\  E {\bf 73}, 036706 (2006).


\bibitem{Sexton:1992nu}
  J.~C.~Sexton and D.~H.~Weingarten,
  Nucl.\ Phys.\  B {\bf 380}, 665 (1992).

\bibitem{Chiu:2011rc}
  T.~W.~Chiu {\it et al.} [ TWQCD Collaboration ],
  PoS {\bf LATTICE2010}, 030 (2010). 
  [arXiv:1101.0423 [hep-lat]], 
  and references therein.

\bibitem{Hasenbusch:2001ne}
  M.~Hasenbusch,
  Phys.\ Lett.\  B {\bf 519}, 177 (2001).

\bibitem{Chiu:HMC}
T.~W.~Chiu et al. [TWQCD Collaboration], 
``Monte Carlo simulation of lattice QCD with the optimal domain-wall fermion", 
in preparation.

\bibitem{Sommer:1993ce} 
  R.~Sommer,
  Nucl.\ Phys.\ B {\bf 411}, 839 (1994)


\bibitem{Chen:2012jy} 
  Y.~C.~Chen, T~.W.~Chiu [TWQCD Collaboration]
  arXiv:1205.6151 [hep-lat].

\bibitem{Colangelo:2005gd}
  G.~Colangelo, S.~Durr and C.~Haefeli,
  Nucl.\ Phys.\  B {\bf 721}, 136 (2005).

\bibitem{Martinelli:1994ty}
  G.~Martinelli, C.~Pittori, C.~T.~Sachrajda, M.~Testa and A.~Vladikas,
  Nucl.\ Phys.\ B {\bf 445}, 81 (1995).

\bibitem{Chiu:2011np}
  T.~W.~Chiu {\it et al.}  [TWQCD Collaboration], 
  ``Nonperturbative renormalization of bilinear operators in lattice QCD 
    with the optimal domain-wall fermion", in preparation.

\bibitem{Noaki:2008iy} 
  J.~Noaki {\it et al.}  [JLQCD and TWQCD Collaboration],
  Phys.\ Rev.\ Lett.\  {\bf 101}, 202004 (2008)

\bibitem{Colangelo:2010et}
  G.~Colangelo, S.~Durr, A.~Juttner, L.~Lellouch, H.~Leutwyler, V.~Lubicz, S.~Necco, C.~T.~Sachrajda {\it et al.},
  Eur.\ Phys.\ J.\  {\bf C71}, 1695 (2011).




\end{thebibliography}
\end{document}